\documentclass[onecolumn,showpacs,floatfix,aps,11pt,nofootinbib]{revtex4}
\usepackage{amssymb,amsmath}
\usepackage[dvips]{graphics,color}
\usepackage{epsfig,yfonts,upgreek}\usepackage{float}
\newcommand{\JJ}{\mathbin{\raisebox{0.25ex}{$\footnotesize
                       \rm\vphantom{I}%
                       \_\hskip -0.25em\_%
                       \vrule width 0.6pt$}}}           

\newcommand{\JJga}{\mathop{\JJ \,}\displaylimits_{g-A}} 
\newcommand{\JJBB}{\mathop{\JJ \,}\displaylimits_B} 
 
\newcommand{\JJAA}{\mathop{\JJ \,}\displaylimits_A} 
\newcommand{\JJgg}{\mathop{\JJ \,}\displaylimits_g} 

\newcommand{\BK}{\mathbb{K}}

\newcommand{\ii}{\textswab{i}}
\newcommand{\jj}{\textswab{j}}
\newcommand{\kk}{\textswab{k}}

\newcommand{\cl}{C \kern -0.1em \ell}     

\newcommand{\ut}[1]{{\setbox0=\hbox{$#1$}\mathsurround=0pt
       \rlap{\raisebox{-0.8\dp0}{\raisebox{-0.8ex}
       {\kern -0.15ex\hbox{$\tiny\sim$}\kern 0.15ex}}}#1}}
\newcommand{\uti}[1]{{\setbox0=\hbox{$#1$}\mathsurround=0pt
       \rlap{\raisebox{-0.8\dp0}{\raisebox{-0.8ex}
       {\kern -0.3ex\hbox{$\tiny\sim$}\kern 0.3ex}}}#1}}

\newdimen\arrayruleHwidth                 
     \makeatletter
     \def\Hline{\noalign{\ifnum0=`}\fi\hrule \@height \arrayruleHwidth
         \futurelet \@tempa\@xhline}
     \makeatother

\newcommand{\beq}{\begin{eqnarray}}
\newcommand{\eeq}{\end{eqnarray}}

\newcommand{\bege}{\begin{equation}}
\newcommand{\enge}{\end{equation}}
\newcommand{\benu}{\begin{enumerate}}
\newcommand{\enu}{\end{enumerate}}

\newcommand{\bbbbox}{\mathop{\Box\kern -5pt\raisebox{.8pt}{$|$}}}

\newcommand{\mt}{\mathcal}

\newcommand{\cle}{{\mt{C}}\ell_{1,3}}

\newcommand{\CC}{\mathbb{C}}

\def\beq{\begin{eqnarray}}
\def\eeq{\end{eqnarray}}

\def\0{\mbox{\boldmath$\displaystyle\mathbf{0}$}}

\def\h00h{\mbox{\boldmath$\displaystyle\mathbf{(1/2,0)\oplus(0,1/2)}$}}

\newcommand{\pu}{\underset{B}{}}

\newcommand{\clb}{\cle^B}

\DeclareMathOperator{\dwedge}{\dot{\wedge}}

\begin{document}

\title{Bilinear Covariants and Spinor Fields Duality  in  Quantum Clifford Algebras}

\author{Rafa\l \;\,Ab\l amowicz}
\email{rablamowicz@tntech.edu}
\affiliation{Department of Mathematics, Box 5054,
Tennessee Technological University, Cookeville, TN 38505, USA
}
\author{Icaro Gon\c calves}\email{icaro.goncalves@ufabc.edu.br}
 \affiliation{
Centro de Matem\'atica, Computa\c c\~ao e Cogni\c c\~ao,
Universidade Federal do ABC, 09210-170, Santo Andr\'e, SP, Brazil}
\affiliation{
Universidade de S\~ao Paulo, Instituto de Matem\'atica e Estat\'istica. 
Rua do Mat\~ao, 1010, 05508-090, S\~ao Paulo, SP,  Brazil}

\author{Rold\~ao da Rocha}
\email{roldao.rocha@ufabc.edu.br} \affiliation{
Centro de Matem\'atica, Computa\c c\~ao e Cogni\c c\~ao,
Universidade Federal do ABC, 09210-170, Santo Andr\'e, SP, Brazil}
\affiliation{International School for Advanced Studies (SISSA), Via Bonomea 265, 34136 Trieste, Italy.}

\begin{abstract}{}
Classification of quantum spinor fields according to quantum bilinear covariants is introduced in a context of quantum Clifford algebras on Minkowski spacetime. Once the bilinear covariants are expressed in terms of algebraic spinor fields, the duality between  spinor and quantum spinor fields is thus  discussed.  Hence, by endowing the underlying spacetime with an arbitrary bilinear form with a antisymmetric part in addition to a symmetric spacetime metric, quantum algebraic spinor fields and deformed bilinear covariants can be constructed. They are therefore compared to the classical (non quantum) ones. Classes of quantum spinor fields are introduced and compared with Lounesto's spinor field classification. A physical interpretation of the deformed parts and the underlying 
$\mathbb{Z}$-grading is proposed.  The existence of an arbitrary bilinear form endowing the spacetime already has been explored in the literature in the context of quantum gravity~\cite{Hawking}. Here, it is shown further to play a prominent role in the structure of Dirac, Weyl, and Majorana spinor fields, besides the most general flagpoles, dipoles  and flag-dipoles ones as well. We introduce a new duality 
between the standard and the quantum spinor fields, by showing 
that when Clifford algebras over vector spaces endowed 
with an arbitrary bilinear form are taken into account, a mixture 
among the classes does occur. Consequently, novel features 
regarding the spinor fields can be derived.
\end{abstract}
\maketitle
\section{Introduction}
The formalism of Clifford algebras is fruitful and advantageous in particular when constructing spinors and Dirac operators, and when introducing periodicity theorems and index theorems. Clifford algebras are essentially associated to a quadratic vector space. Notwithstanding, the underlying vector space may be endowed with  bilinear form that is not necessarily symmetric~\cite{chevalley}. For instance, symplectic Clifford algebras are objects of huge interest. More generally, when the underlying vector space is endowed with an arbitrary bilinear form, it exhibits prominent features, especially in the area of algebra representation. The most evident feature distinguishing the so called \textit{quantum} Clifford algebras (of an arbitrary form) from \textit{orthogonal} Clifford algebras (of a quadratic form) is that a different $\mathbb{Z}$-grading arises, despite of the  $\mathbb{Z}_2$-grading being the same.

The most general Clifford algebras of multivectors \cite{Oziewicz97} are further named quantum Clifford algebras. The arbitrary bilinear form of a quantum Clifford algebra defines a $\mathbb{Z}$-grading which is required in physics. For instance, it is employed in  quantum mechanical setups. In fact, when one analyzes functional hierarchy equations in quantum field theory (QFT),  Clifford algebras are applied to provide a description of these functionals. At least time-ordering and normal-ordering are needed in QFT~\cite{Fsr1}, and singularities due to the reordering procedures  are no longer present~\cite{Fsr2}, in this framework. In a very particular case~\cite{Fsr3}, quantum Clifford algebras lead to Hecke algebras, which play a major role, for instance, in the discussion of Yang-Baxter equation and link invariants, knots,  and integrable systems in statistical physics. Moreover, the coalgebraic and the Hopf algebraic structure associated to those algebras can be thoroughly defined \cite{ozie,conformal,ber1}. More strictly, a Clifford algebra is well known not to be a Hopf algebra, as it was demonstrated in \cite{rota}. However,  invariants have to be different of zero, in order to manufacture a cobilinear form and a antipodal Clifford convolution, which effectively acts as a Clifford Hopf algebra \cite{gebra}. Otherwise, further extensions are requested in order to accomplish this achievement. In particular, there is an extension to infinite supersymmetric Hilbert spaces, that can emulate a Hopf algebra~\cite{rota}.

The isomorphism between Clifford algebras $\cl(V,g)$ constructed  over a vector space~$V$ endowed with a symmetric bilinear form~$g$, and quantum Clifford algebras $\cl(V,B)$ --- built over~$V$ equipped with an arbitrary bilinear form~$B$~\cite{vacua} --- is here further surveyed. The algebra $\cl(V,B)$,  besides being isomorphic to $\cl(V,g)$, is indeed different in what concerns its vacuum structure, the structure of its representations, and dualities, as comprehensively discussed in ~\cite{Fsr1, Fsr2, Fsr3,gebra,vacua}. 
  
From a more physical point of view, references can be found for a non symmetric model for gravity, and some applications. For instance, the rotation curves of galaxies and cosmology, without adducing dominant dark matter and identifying dark energy with the cosmological constant, can be obtained~\cite{moffat,moffat01,moffat1,moffat2,moffat3,moffat4} by just considering the spacetime metric to have symmetric \emph{and}  antisymmetric parts. Moreover, a gravitational theory based on general relativity, that depends on non symmetric tensors playing an important role in the spacetime metric, was formulated and discussed in~\cite{moffat,moffat1,hammond}. Unexpected properties on thick brane-world scenarios have been achieved in this framework \cite{Ghosh:2012jh}.

Different  classes of spinor fields have been had unexpected physical applications, even for new possibilities of fermions of mass dimension one in Minkowski spacetime.
 Precursory results have applications in various aspects of field theory \cite{alex,horvath,2cylee,cyleeI,canj} and comprise cosmology  \cite{dasi,shankara,coincidence,liu,Annalen Phys.,vign,lucafabbri}, thick branes gravity \cite{Ghosh:2012jh} and black hole thermodynamics \cite{elkoh} as well. 
Phenomenological aspects of recently new  found particles, described by non-standard spinor fields, has been proposed  \cite{accele,alve}. For instance, in the background of  $f(R)$ modified and conformal gravities in a framework with torsion, the Dirac equation discloses solutions that \emph{are not} Dirac spinor fields. Instead, they have been shown to be the first physical particle manifestation of flag-dipoles spinor fields \cite{riemanncartan}. Moreover, to depart from the standard Lorentzian spacetime structure conveys novel applications in cosmology and condensed matter, as the results in \cite{alex,wess,Asselmeyer:1995jm,exotic} evince that non-connected spacetimes present surprising properties of non-standard spinor fields. Recently all the  spinor classes have been characterized thoroughly \cite{Cavalcanti:2014wia}.

The goal of this paper is to provide a complete classification of quantum algebraic spinor fields in the context of quantum Clifford algebras and associate to them bilinear covariants, constructing an important duality between the spinor fields in Lounesto's classification and their quantum counterparts. It potentially shall provide 
the well known results derived from the Lounesto spinor field classification in the regime of quantum gravity. 
The paper is organized as follows: in Section~II, Clifford algebras, bilinear covariants and associated spinor field classification are revisited. In Section~III, the correspondence between classical and algebraic spinor fields is obtained. In Section~IV, quantum algebraic spinor fields and their important properties are introduced and scrutinized. In Section~V, the spinor field classification, according to their bilinear covariants, is described in the quantum Clifford algebraic formalism.  The duality between spinor fields and quantum spinor fields is provided and summarized in Table~I. In Section~VI, we digress about possible further developments. Appendix~A contains detailed computations which are reported in Section~IV.

\section{Bilinear Covariants}
\label{w2}

Given an arbitrary element $\psi$ in the exterior algebra $\bigwedge (V)$ of an $n$-dimensional real vector space $V$, the 
 reversion is provided by $\tilde{\psi}=(-1)^{[k/2]}\psi$ where $[k]$ stands for the integer part of $k$. By endowing  $V$ with a symmetric bilinear form 
${g}:V\times V\rightarrow \mathbb{R}$, the extension of ${g}$ to $\bigwedge(V)$ can be regarded, and given $\xi,\kappa,\phi \in \bigwedge (V)$, the {left contraction} satisfy  ${g}(\kappa \lrcorner \phi, \xi)={g}(\phi ,\tilde{\kappa}\wedge \xi )$.  The Clifford product between ${w}\in V$ and $\kappa \in \bigwedge (V)$ is denoted by juxtaposition and prescribed by  
${{w}}\kappa ={{w}}\wedge \kappa +{{w}}\lrcorner \kappa $. The 2-uple $(\bigwedge (V),{g})$ endowed with the Clifford product is denoted either by  
$\cl _{p,q}$ or $\cl (V,{g})$ equivalently, and named the Clifford algebra of $(V,g)\cong \mathbb{R}^{p,q}$, where $p+q=n$ is related to the signature of spacetime, being $p$ the number of positive signs in the diagonal  metric. We shall use throughout the text the Einstein's summation convention.

In order to revisit the bilinear covariants,  all spinor fields hereupon reside in Minkowski spacetime~$M\cong\mathbb{R}^{1,3}$ with metric components $\eta _{\mu \nu }=\mathrm{diag}(1,-1,-1,-1)$. The set $\{{\rm e}_{\mu }\}$ stands for sections of the frame bundle ${P}_{\mathrm{SO}_{1,3}^{e}}(M)$~\cite{moro} with  dual 
basis $\{{\theta}^{\mu }\}$. Classical spinor fields carrying a ${(1/2,0)}\oplus {(0,1/2)}$ representation of the Lorentz group component connected to the identity $\mathrm{SL}(2,\mathbb{C)}\simeq\mathrm{Spin}_{1,3}^{0}$ are sections of the vector bundle ${P}_{\mathrm{Spin}_{1,3}^{0}}(M)\times _{\rho }\mathbb{C}^{4},$ where $\rho$ stands for the ${(1/2,0)}\oplus {(0,1/2)}$ representation of 
$\mathrm{Spin}_{1,3}^{0}$ in the vector space $\mathbb{C}^{4}$. 

Let us thus consider a spinor field $\psi \in \sec {P}_{\mathrm{Spin}_{1,3}^{0}}(M)\times _{\rho }\mathbb{C}^{4}$ and the bilinear covariants, that are   sections of the exterior 
bundle~$\bigwedge(TM)$ \cite{yvon,taka,fierz,lou2} (hereupon $\iota$ shall denote the imaginary unit):
\begin{align}
\upsigma & =\psi ^{\dagger }\upgamma _{0}\psi , \notag \\ \mathbf{J}&=J_{\mu }\theta ^{\mu }=\psi ^{\dagger }\upgamma _{0}\upgamma _{\mu }\psi \theta ^{\mu }, \notag \\ 
\mathbf{S}&=S_{\mu \nu }\theta ^{\mu}\wedge\theta^{ \nu }=\frac{1}{2}\psi ^{\dagger }\upgamma _{0}i\upgamma _{\mu \nu }\psi \theta ^{\mu }\wedge \theta ^{\nu },  \notag \\
\mathbf{K}& =K_{\mu }\theta ^{\mu }=\psi ^{\dagger }\upgamma _{0}\iota\upgamma _{0}\upgamma_{1}\upgamma_{2}\upgamma_{3}\upgamma _{\mu }\psi \theta ^{\mu }, \notag \\ \upomega  &=-\psi ^{\dagger }\upgamma _{0}\upgamma_{0}\upgamma_{1}\upgamma_{2}\upgamma_{3}\psi .  
\label{fierz}
\end{align}
For $\mu ,\nu
,\tau =0,1,2,3$, the set $\{\mathbb{I},\upgamma _{\mu },\upgamma _{\mu }\upgamma _{\nu },\upgamma _{\mu }\upgamma _{\nu }\upgamma _{\tau},\upgamma _{0}\upgamma _{1}\upgamma _{2}\upgamma _{3}\}$,  for $\mu <\nu <\tau$,   is a basis for the space of matrices $4\times 4$ over the complex field, and $\upgamma _{\mu }\upgamma _{\nu }+\upgamma _{\nu }\upgamma _{\mu }=2\eta _{\mu \nu }\mathbb{I}$, hence constituting  a Clifford algebra for $M$. 
Classically, $\mathbf{J}$ stands for  the time-like  current of probability,  the space-like covector $\mathbf{K}$ regards the direction of the electron spin, and 
$\mathbf{S}$ provides the distribution of intrinsic angular momentum. The bilinear covariants satisfy the  Fierz identities \cite{moro,cra,lou2,hol}:
\begin{equation}
(\upomega +\upsigma\upgamma_{0}\upgamma_{1}\upgamma_{2}\upgamma_{3})\mathbf{S}=\mathbf{K}\wedge\mathbf{J},\qquad \mathbf{J}^{2}=\upomega ^{2}+\upsigma^{2} = - \mathbf{K}^{2},\qquad\mathbf{K}\cdot\mathbf{J}=0\,.  \label{fi}
\end{equation}
\noindent 
When $\upomega =0=\upsigma$, the spinor field is said to be {singular}.
 Lounesto's spinor field classification is based on six disjoint  spinor field classes~\cite{lou2}, where $\mathbf{J}, \mathbf{K}, \mathbf{S}\neq 0$ for spinor fields that \emph{are not} singular:
\begin{itemize}
\item[1)] $\upsigma\neq0,\;\;\; \upomega \neq0$\qquad\qquad\qquad\qquad\qquad4) $\upsigma= 0 = \upomega , \;\;\;\mathbf{S}\neq 0, \;\;\;\mathbf{K}\neq0$%
\label{Elko11}
\item[2)] $\upsigma\neq0,\;\;\; \upomega = 0$\label{dirac1}\qquad\qquad\qquad\qquad\qquad5) $\upsigma= 0 = \upomega , \;\;\;\mathbf{S}\neq0,\;\;\; \mathbf{K}=0$%
\label{tipo41}
\item[3)] $\upsigma= 0, \;\;\;\upomega \neq0$\label{dirac21} \qquad\qquad\qquad\qquad\qquad\!6) $\upsigma= 0 = \upomega , \;\;\; \mathbf{S}=0, \;\;\; \mathbf{K} \neq 0$%
\end{itemize}
\noindent 
The first three types of spinor fields are named Dirac spinor fields, whilst the last three are known, respectively, as flag-dipole, flagpole and dipole spinor fields. Flagpole spinor fields have been recently explored as candidates for dark matter, see, e.g., ~\cite{horvath,2cylee,cyleeI}, and range across Elko and Majorana spinor fields. Flag-dipole ones are pictured for instance by recently new found particles that are solutions of the Dirac equation in Einstein-Sciama-Kibble gravities \cite{riemanncartan}. Dipole spinor fields have Weyl spinor fields as their most known representative.  

The multivector field $\mathcal{Z}=\upsigma+\mathbf{J}+\iota\mathbf{S}+\iota\mathbf{K}\upgamma_{0}\upgamma_{1}\upgamma_{2}\upgamma_{3}+\upomega \upgamma_{0}\upgamma_{1}\upgamma_{2}\upgamma_{3}$ is called a {Fierz aggregate} provided that $\upsigma,\mathbf{J},\mathbf{S},\mathbf{K},\upomega$ satisfy the Fierz identities \cite{lou2}. Moreover, $\mathcal{Z}$ is named  a {boomerang} when $\upgamma _{0}\mathcal{Z}^{\dagger}\upgamma_{0}=\mathcal{Z}$. For {singular} spinor fields, all the Fierz identities in  (\ref{fi}) are replaced by \cite{cra,lou2}: 
\begin{alignat}{2}
\mathcal{Z}^{2} &=4\upsigma \mathcal{Z},\notag\\ 
\mathcal{Z}\upgamma_{\mu}\mathcal{Z} &=4J_{\mu}\mathcal{Z},\notag\\  \mathcal{Z}\upgamma_{\mu}\upgamma_{\nu}\mathcal{Z}&=-4\iota S_{\mu\nu}\mathcal{Z},\notag\\ \mathcal{Z}\upgamma_{0}\upgamma_{1}\upgamma_{2}\upgamma_{3}\upgamma_{\mu}\mathcal{Z}&=-4\iota K_{\mu}\mathcal{Z},\notag\\
\mathcal{Z}\upgamma_{0}\upgamma_{1}\upgamma_{2}\upgamma_{3}\mathcal{Z}&=-4\upomega  \mathcal{Z}.\nonumber
\end{alignat}

\section{Classical, Algebraic Spinors and Spinor Operators}
\label{scle} 
In this section we briefly revisit and further develop the formalism of algebraic spinor fields in order to make it easier to define quantum spinor fields in later sections. An arbitrary element of ${\mathcal{C}}\ell _{1,3}$ is written as:
\begin{gather}
\epsilon = h+h^{\mu}{\rm e}_{\mu}+h^{\mu\nu}{\rm e}_{\mu}{\rm e}_{\nu}+h^{\mu\nu\tau}{\rm e}_{\mu}{\rm e}_{\nu}{\rm e}_\tau+p{\rm e}_0{\rm e}_1{\rm e}_2{\rm e}_3. 
\label{spinor}
\end{gather}
Given an isomorphism ${\mathcal{C}}\ell _{1,3}\simeq {\mathcal{M}}(2,\mathbb{H})$, in order to obtain a spinor representation of ${\mathcal{C}}\ell _{1,3}$, a primitive idempotent 
$f=\frac{1}{2}(1+{\rm e}_{0})$ defines a minimal left ideal ${\mathcal{C}}\ell _{1,3}f \subset{\mathcal{C}}\ell _{1,3}$. A general element in this ideal has thus the structure 
\begin{equation}
 \Omega =(b^{1}+b^{2}{\rm e}_{2}{\rm e}_{3}+b^{3}{\rm e}_3{\rm e}_{1}+b^{4}{\rm e}_{1}{\rm e}_{2})f+(b^{5}+b^{6}{\rm e}_{2}{\rm e}_{3}+b^{7}{\rm e}_{3}{\rm e}_{1}+b^{8}{\rm e}_{1}{\rm e}_{2}){\rm e}_0{\rm e}_1{\rm e}_2{\rm e}_3\,f\,.
\label{kkk}
\end{equation}
\noindent 
If  $\Omega = \epsilon f \in \cl_{1,3}f$ is set, hence the components in (\ref{spinor}) and (\ref{kkk}) are precisely related by:
\begin{eqnarray}
b^{1} &=&h+h^{0},\qquad b^{2}=h^{023}+h^{23},\qquad b^{3}=-h^{13}-h^{013},\qquad b^{4}=h^{012}+h^{12},\nonumber\\
b^{5} &=& p-h^{123},\qquad b^{6}=h^{1}-h^{01},\qquad b^{7}=h^{2}-h^{02},\qquad b^{8}=h^{3}-h^{03}.\nonumber
\end{eqnarray}
\noindent 
By assigning $\textswab{i}={\rm e}_{2}{\rm e}_{3},\;\textswab{j}={\rm e}_{3}{\rm e}_{1},$ and $\textswab{k}={\rm e}_{1}{\rm e}_{2}$,  the set $\{1, \textswab{i},\textswab{j},\textswab{k}\}$ provides a basis for an algebra that is isomorphic to the quaternion algebra~$\mathbb{H}$.  The minimal left ideal $\cl_{1,3}f$ is also a right $\mathbb{H}$-module: thereby the quaternionic coefficients should be written to the {right} of $f$. The two quaternions appearing as coefficients in~(\ref{kkk}), namely, 
\begin{eqnarray}
q_1 &=& b^1+b^2{\rm e}_{2}{\rm e}_{3}+b^3{\rm e}_{3}{\rm e}_{1}+b^4{\rm e}_{1}{\rm e}_{2}\in\mathbb{K},\\\notag q_2 &=& b^5+b^6{\rm e}_{2}{\rm e}_{3}+b^7{\rm e}_{3}{\rm e}_{1}+b^8{\rm e}_{1}{\rm e}_{2}\in\mathbb{K}\,,
\end{eqnarray}
where $\mathbb{K}=f\cl_{1,3}f = {\rm span}_\mathbb{R}\{1, {\rm e}_{2}{\rm e}_{3},{\rm e}_{3}{\rm e}_{1},{\rm e}_{1}{\rm e}_{2}\} \cong \mathbb{H}$,  commute with both the elements $f$ and ${\rm e}_{0}{\rm e}_1{\rm e}_2{\rm e}_{3}$. Hence, 
$
fq_1 + {\rm e}_0{\rm e}_1{\rm e}_2{\rm e}_3fq_2 = q_1f + q_2{\rm e}_0{\rm e}_1{\rm e}_2{\rm e}_3f, 
$ 
and the left ideal $\cl_{1,3}f$ is a right module over $\BK$ with a basis $\{f,{\rm e}_0{\rm e}_1{\rm e}_2{\rm e}_3\, 
f\}$.

Matrices representing the  basis elements $\{{\rm e}_{\mu }\}$ can be obtained: 
\begin{equation}
{\rm e}_{1}=%
{\scriptstyle\begin{pmatrix}
0 & \textswab{i} \\ 
\textswab{i} & 0%
\end{pmatrix}}%
,\quad{\rm e}_{2}=%
{\scriptstyle\begin{pmatrix}
0 & \textswab{j} \\ 
\textswab{j} & 0%
\end{pmatrix}}%
,\quad{\rm e}_{3}=%
{\scriptstyle\begin{pmatrix}
0 & \textswab{k} \\ 
\textswab{k} & 0%
\end{pmatrix}}\,,\quad{\rm e}_{0}=%
{\scriptstyle\begin{pmatrix}
1 & 0 \\ 
0 & -1%
\end{pmatrix}}%
\end{equation}%
\noindent 
as well as matrices representing elements $f$ and ${\rm e}_0{\rm e}_1{\rm e}_2{\rm e}_3f$, namely, 
$$
[f]={\scriptstyle\begin{pmatrix}
1 & 0 \\ 
0 & 0%
\end{pmatrix}} \quad \mbox{and} \quad 
[{\rm e}_0{\rm e}_1{\rm e}_2{\rm e}_3f]=%
{\scriptstyle\begin{pmatrix}
0 & 0 \\ 
1 & 0%
\end{pmatrix}}.
$$ 
Thus, the general element $\mathbf{\epsilon}\in{\mathcal{C}}\ell_{1,3}$ in (\ref{spinor}) has the matrix representation:
\begin{eqnarray}\label{quat}
\mathbf{\epsilon} = {\scriptstyle\begin{pmatrix}q_1 & q_2 \\ 
q_3 & q_4%
\end{pmatrix}}\,,\end{eqnarray}
where
\begin{eqnarray}
q_1&=& 
h+ h^0 + (h^{23} + h^{023})\textswab{i}  
-(h^{13} + h^{013})\textswab{j} + (h^{12} + h^{012})\textswab{k} \nonumber\\
q_2&=&
-(h^{123} + p) + (h^1 + h^{01})\textswab{i} +  
(h^2 + h^{02})\textswab{j} + (h^3 + h^{03})\textswab{k}\nonumber\\ 
q_3&=&(p-h^{123}) + (h^1 - h^{01})\textswab{i} 
+(h^2 - h^{02})\textswab{j} + (h^3 - h^{03})\textswab{k}\nonumber\\
q_4&=& (h- h^0) + (h^{23} - h^{023})\textswab{i} +  
(h^{013}-h^{13})\textswab{j} + (h^{12} - h^{012})\textswab{k}\label{repoo}
\end{eqnarray}
Spinor fields  have been similarly represented by differential forms since the works by  Landau, Fock, and Ivanenko  (in 1928), with a large amount of important applications in diverse areas of physics and mathematics~\cite{lou2}. An even element 
$\Psi \in \cl_{1,3}^{+}$ (hereupon  $\cl_{1,3}^{+}$ is used to denote the even subalgebra of $\cl_{1,3}$) corresponding to the so called {spinor operator} can be clearly written
as 
\begin{equation}
\Psi=h+h^{\mu\nu}{\rm e}_{\mu}{\rm e}_{\nu}+p{\rm e}_{0}{\rm e}_1{\rm e}_2{\rm e}_{3}\,,
\label{400}
\end{equation}%
\noindent which in the light of the quaternionic representation in Eqs.~(\ref{quat}) and (\ref{repoo}) is provided by 
\begin{equation}
{\scriptstyle\begin{pmatrix}
q_{1} & -q_{2} \\ 
q_{2} & q_{1}%
\end{pmatrix}}%
= 
\begin{pmatrix}
h+h^{23}\textswab{i}-h^{13}\textswab{j}+h^{12}\textswab{k} & \quad
-p+h^{01}\textswab{i}+h^{02}\textswab{j}+h^{03}\textswab{k} \\ 
p-h^{01}\textswab{i}-h^{02}\textswab{j}-h^{03}\textswab{k} & \quad
h+h^{23}\textswab{i}-h^{13}\textswab{j}+h^{12}\textswab{k}
\end{pmatrix} . 
\end{equation}%
The vector space isomorphisms $ \mathbb{C}^{4} \;(\simeq \mathbb{H}^{2}) \simeq  {\mathcal{C}}\ell _{1,3}^{+} \simeq {\mathcal{C}}\ell _{1,3}\frac{1}{2}(1+{\rm e}_{0})$  show respectively the
equivalence among the classical, the operatorial, and the algebraic definitions of a spinor. The spinor space $\mathbb{H}^{2}$, carrying the 
${(1/2,0)}\oplus {(0,1/2)}$ or ${(1/2,0)}$, or ${(0,1/2)}$ representations of $\mathrm{SL}(2,\mathbb{C)}$, is hence isomorphic to the minimal left ideal 
${\mathcal{C}}\ell _{1,3}\frac{1}{2}(1+{\rm e}_{0})$ (corresponding to the algebraic spinor) and  isomorphic to the even subalgebra ${\mathcal{C}}\ell _{1,3}^{+}$  (corresponding to
the spinor operator) likewise. The Dirac spinor field is expressed accordingly as 
\begin{equation}
{\scriptstyle\begin{pmatrix}
q_{1} & -q_{2} \\ 
q_{2} & q_{1}%
\end{pmatrix}} [f]
= {\scriptstyle\begin{pmatrix} q_1 & 0\\q_2 & 0 \end{pmatrix}}\cong{\scriptstyle\begin{pmatrix} q_1\\q_2 \end{pmatrix}} =\left( 
\begin{array}{c}
h+h^{23}\textswab{i}-h^{13}\textswab{j}+h^{12}\textswab{k} \\ 
p-h^{01}\textswab{i}-h^{02}\textswab{j}-h^{03}\textswab{k}%
\end{array}%
\right) \in \cl_{1,3}f\simeq \mathbb{H}\oplus \mathbb{H}.  \label{hh}
\end{equation}%
\noindent 
Returning to Eq.~(\ref{400}), and using for instance the standard representation 
\begin{gather}
1   \mapsto {\scriptstyle\begin{pmatrix}  1 & 0\\ 0 & 1\end{pmatrix}}, \quad
\ii \mapsto {\scriptstyle\begin{pmatrix}  \iota & \phantom{-}0\\ 0 & -\iota \end{pmatrix}}, \quad
\jj \mapsto {\scriptstyle\begin{pmatrix}  \phantom{-}0 & 1\\ -1 & 0 \end{pmatrix}}, \quad
\kk \mapsto {\scriptstyle\begin{pmatrix}  0 & \iota\\ \iota & 0 \end{pmatrix}}\,, 
\label{quatrepr}
\end{gather}
the complex matrix associated to the spinor operator $\Psi$ in~(\ref{400})  can be written as 
\begin{equation*}
\begin{pmatrix} 
h+h^{23}\iota & -h^{13}+h^{12}\iota & -p+h^{01}\iota & h^{02}+h^{03}\iota \\
h^{13}+h^{12}\iota & h-h^{23}\iota & -h^{02}+h^{03}\iota & -p-h^{01}\iota\\
p-h^{01}\iota & -h^{02}-h^{03}\iota & h+h^{23}\iota & -h^{13}+h^{12}\iota\\
h^{02}-h^{03}\iota & p+h^{01}\iota & h^{13}+h^{12}\iota & h-h^{23}\iota\\
\end{pmatrix} 
:=%
\begin{pmatrix} 
\upphi_1 & -\upphi_2^{*} & -\upphi_3 &  \phantom{-}\upphi_4^{*}\\
\upphi_2 & \phantom{-}\upphi_1^{*} & -\upphi_4 & -\upphi_3^{*}\\
\upphi_3 & -\upphi_4^{*} & \phantom{-}\upphi_1 & -\upphi_2^{*}\\
\upphi_4 & \phantom{-}\upphi_3^{*} & \phantom{-}\upphi_2 & \phantom{-}\upphi_1^{*}
\end{pmatrix}.
\end{equation*}

The Dirac spinor $\psi $ is an element of the minimal left ideal $(\mathbb{C}\otimes \cl _{1,3})f$ generated by $ f=\frac{1}{4}(1+{\rm e}_{0})(1+\iota {\rm e}_{1}{\rm e}_{2})$, fixing the Dirac  representation that sends the basis vectors ${\rm e}_{\mu }$ to $\upgamma _{\mu }\in \mathrm{End}(\mathbb{C}^{4}$).  Hence, $\psi =\Omega \frac{1}{2}(1+\iota\upgamma _{1}\upgamma_{2})\in (\mathbb{C}\otimes \cl _{1,3})f,$ where 
$2\Omega$ is the real part of $\psi$. It follows that the algebraic spinor 
\begin{equation}
\psi \sim  
\begin{pmatrix}
\upphi_1 & 0 & 0 & 0 \\ 
\upphi_2 & 0 & 0 & 0 \\ 
\upphi_3 & 0 & 0 & 0 \\ 
\upphi_4 & 0 & 0 & 0%
\end{pmatrix}\in (\CC\otimes \cl _{1,3})f
\end{equation}
is identified to the usual form corresponding to a classical spinor: 
\begin{equation}
\begin{pmatrix}
\upphi_1 \\ 
\upphi_2 \\ 
\upphi_3 \\ 
\upphi_4%
\end{pmatrix}%
\equiv
\begin{pmatrix}
\psi_1 \\ 
\psi_2 \\ 
\psi_3 \\ 
\psi_4%
\end{pmatrix}%
\in \mathbb{C}^{4}\,.  \label{c4}
\end{equation} This provides an immediate identification between the algebraic and the classical definitions of a spinor field.

\section{Quantum Algebraic Spinor Fields}
We let $\cl(V,g)$ denote the Clifford algebra over a vector space $V$ endowed with a symmetric bilinear form $g:V\times V\rightarrow\mathbb{R}$ that defines a quadratic form $Q:V\rightarrow\mathbb{R}$, and we let $\cl(V,B)$ denote the quantum Clifford algebra  constructed over~$V$ equipped with an arbitrary bilinear form~$B=g+A$ where $A:V\times V\rightarrow\mathbb{R}$ denotes hereon the antisymmetric part of $B$. Such two algebras are isomorphic as $\mathbb{Z}_2$-graded algebras~\cite{vacua}. 
Moreover, as $Q(u) = B(u, u)$, the Clifford algebra $\cl(V,g)$ is isomorphic to the associative algebra defined on the exterior algebra $\bigwedge(V)$ with (associative) product determined by
\beq
u\pu \psi := u \wedge \psi + u\JJBB \psi\label{b11}\eeq
for any $u \in V$ and $\psi \in \bigwedge(V)$, 
where $\displaystyle{u\JJBB \psi} = \delta_u^B(\psi)$, is the antiderivation of degree $-1$ on the exterior algebra $\bigwedge(V)$ determined by $u\displaystyle{\JJBB} v = B(u, v)1$,  for any $v \in V$.
The interesting fact of this result by Chevalley is that this is independent of $B$. Hence, the $B$-product determined by Eq.(\ref{b11}) is isomorphic to the usual product in $\cl(V,g)$ (where $A = 0$). Nevertheless, it shall be clear in what follows that  although 
$\cl(V,B)$ and $\cl(V,g)$ are isomorphic, 
novel signatures can be evinced when 
we look at the Clifford algebra $\cl(V,B)$ from the point of view of $\cl(V,g)$.

The antisymmetric part $A$ of the arbitrary bilinear form $B$ have already been analyzed and interpreted \cite{vacua}. This transparent analysis compares Clifford methods to $q$-deformed counter parts, and scrutinizes the vacuum structure of such theories likewise, including BCS superconductivity and the gap equation. Moreover, if we denote the entries of~$B$ by $B_{ij}$, the symmetric part $g_{ij} = (B_{ij}+B_{ji})/2$ and the antisymmetric part $A_{ij} = (B_{ij}-B_{ji})/2$, the parameters $A_{ij}$ may describe environmental  settings, like for instance, the external temperature,  which induce an interaction mediated by the~$A_{ij}$'s. The procedure of $q$-deformation is related to such interactions as the $R$-matrix described the scattering matrix,
in that case form Cooper pairs~\cite{hecke,arbi}. It is very interesting to note moreover that the $\bmod\, 8$ periodicity structure fails to hold in $\cl(V,B)$,  as  an exotic indecomposable 8-dimensional spinor state is achieved~\cite{arbi}. 

The quantum Clifford algebra $\cl(V,B)$ has two alternative underlying vector space structures: the usual exterior algebra $\bigwedge (V)$ with the standard wedge product and the dotted-exterior algebra $\dot{\bigwedge}(V)$  with a bilinear associative multiplication  $\dot\wedge$ defined in terms of the standard exterior product.

In order to define this multiplication, consider the map $V \rightarrow {\rm End}(\bigwedge(V))$ such that $u\mapsto l_u +\delta_u^A$, where  $l_u(\psi) = u \wedge \psi$ for any $\psi\in\bigwedge(V)$, and $\delta_u^A$ is the antiderivation of $\bigwedge(V)$ such that
$\delta_u^A(v) = A(u, v)1$ for any $u, v \in V$.
 Since ($l_u + \delta_u^A)^2 = 0$, this map induces a homomorphism $\lambda: \bigwedge(V) \rightarrow {\rm End}(\bigwedge(V))$.  For any $u_1,\ldots,u_r\in V$, 
\beq
\lambda(u_1{\wedge}\cdots {\wedge} u_r)(1) &:=&(l_{u_1} +\delta_{u_1}^A)\circ\cdots\circ (l_{u_{r}} +\delta_{u_{r}}^A)(1)\nonumber\\
&=& u_1\wedge\cdots \wedge u_r\; + \;\;\text{terms in}\;\; \bigoplus_{i\geq 0}\bigwedge^{r-2i}(V).
\eeq
This shows that the linear map $\varphi: \bigwedge(V) \rightarrow \bigwedge(V)$ given by $\varphi(\psi) = \lambda(\psi)(1)$, for any $\psi\in\bigwedge(V)$, is bijective. Moreover, $\varphi(1) = 1$ and $\varphi(u)=u$ for any $u\in V$. Also we have
\beq
\varphi(u\wedge v)&=& (l_u+\delta_u^A)\circ(l_v+\delta_v^A)(1) = (l_u+\delta_u^A)(v)\nonumber\\&=&u\wedge v + A(u,v)1\,,
\eeq\nonumber for any $u,v\in V$, and similarly 
\beq
\varphi(u\wedge v\wedge w)&=& (l_u+\delta_u^A)\circ(l_v+\delta_v^A)\circ(l_w+\delta_w^A)(1)\nonumber\\ 
&=& (l_u+\delta_u^A)(v\wedge w + A(v,w)1)\nonumber\\&=&u\wedge v\wedge w + A(u,v)w + A(v,w)u+A(w,u)v\,,
\eeq\nonumber for any $u,v,w\in V$.

Define the new associative multiplication
 $\dot\wedge$ on $\bigwedge(V)$ by the formula
\beq
\psi \dwedge \xi = \varphi\left(\varphi^{-1}(\psi) \wedge \varphi^{-1}(\xi)\right),
\eeq\noindent 
for any $\psi,\xi \in \bigwedge(V)$ and denote by $\dot\bigwedge(V)$ the algebra defined on $\bigwedge(V)$ with this new multiplication. From its definition, it is clear that $\varphi$ is an isomorphism between $\bigwedge(V)$ and  
$\dot\bigwedge(V)$, hence the new product is associative.
Thus the above equations give 
\beq
u_1\dot{\wedge}\cdots \dot{\wedge} u_r = \varphi(u_1)\dwedge\cdots\dwedge\varphi(u_r) = \varphi(u_1 \wedge \cdots\wedge u_r)\,.\eeq  
Thus, in particular, 
\begin{eqnarray}
u\dwedge v&=&u\wedge v + A(u,v)1\,,\label{deff}\\
\label{dott}
u\dwedge v\dwedge w &=& u\wedge v\wedge w +  A(u,v)w + A(w,u)v + A(v,w)u\,.
\end{eqnarray} Hence we see that $\dot{\bigwedge}^{2}(V)\subset{\bigwedge}^{0}(V)\oplus{\bigwedge}^{2}(V)$ and $\dot{\bigwedge}^{3}(V)\subset{\bigwedge}^{1}(V)\oplus{\bigwedge}^{3}(V)$. In other words, the 3-forms in the dotted-exterior algebra $\dot{\bigwedge}(V)$ are indeed related to an element which is  the sum of 1-forms and 3-forms in the standard exterior algebra $\bigwedge(V)$. 

The algebra $\dot{\bigwedge}(V)$ is hence $\mathbb{Z}$-graded: $\dot{\bigwedge}(V) = \bigoplus_{r= 0}^n \dot{\bigwedge}^r(V)$, where $\dot{\bigwedge}^r(V)=\varphi\left({\bigwedge}^r(V)\right)$ is the linear span of the elements $u_1\dwedge\cdots\dwedge u_r$. 
Each homogeneous subspace in the above direct sum is a proper subset of the direct sum of subspaces in the standard exterior algebra that have the same $\mathbb{Z}_2$-grading: 
\begin{gather}
\dot{\bigwedge}^{2i}(V)\subset{\bigwedge}^0(V)\oplus{\bigwedge}^2(V)\oplus\cdots=\bigoplus_{j=0}^{i}{\bigwedge}^{2j}(V),\notag\\
\dot{\bigwedge}^{2i+1}(V)\subset{\bigwedge}^1(V)\oplus{\bigwedge}^3(V)\oplus\cdots=\bigoplus_{j=1}^{i} {\bigwedge}^{2j+1}(V).
\end{gather}
It is straightforward to realize that $\dot\bigwedge(V)$ and $\bigwedge(V)$ have different $\mathbb{Z}$-gradings although they present the same $\mathbb{Z}_2$-grading. 
Moreover, $$\bigoplus_{i=0}^r\dot{\bigwedge}^{i}(V)=\bigoplus_{i=0}^r{\bigwedge}^{i}(V),\;\; \forall\, r = 0,\ldots, n,$$
and this is true if only either even or odd indexes are used likewise.

Therefore, the elements of $\dot\bigwedge^k(V)$ behave like $k$-forms \emph{with respect to the algebra} $\dot\bigwedge(V)$ but they are inhomogeneous sums of $k$-forms, ($k$-$2$)-forms, ($k$-$4$)-forms and thereon, in general, with respect to the exterior algebra $\bigwedge(V)$. Which exterior algebra -- either the dotted or the undotted one -- we shall use to describe physics, has been a choice heretofore. 

Instead, when $V = \mathbb{R}^{1,3}$, experimentalists should measure the $B$-bilinear covariants, which will be defined in the next section, and use the undotted basis 
$$
\mathcal{B} = \{1, {\rm e}_\mu, {\rm e}_\mu\wedge {\rm e}_\nu, {\rm e}_\mu\wedge {\rm e}_\nu\wedge {\rm e}_\tau, {\rm e}_{0}\wedge {\rm e}_1 \wedge {\rm e}_2 \wedge {\rm e}_3\} \subset \bigwedge(V),
$$
and not the dotted basis 
$$
\dot{\mathcal B}=\{1, {\rm e}_\mu, {\rm e}_\mu\dwedge {\rm e}_\nu, {\rm e}_\mu\dwedge {\rm e}_\nu\dwedge {\rm e}_\tau, {\rm e}_{0}\dwedge {\rm e}_1\dwedge {\rm e}_2 \dwedge {\rm e}_3\}\subset\dot\bigwedge(V),
$$
in order to perform measurements. This has a very strong geometric reason, which might prefer one over the other basis by considering an independent coordinatization of point and hyperplane spaces
in a projective setting
~\cite{vacua}. 

Although the exterior algebras $\bigwedge(V)$ and $\dot\bigwedge(V)$  are isomorphic (indeed equal) as vector spaces, having the same $\mathbb{Z}_2$-grading, reversion, as opposed to graded involution which depends upon the invariant $\mathbb{Z}_2$-grading, depends on the chosen antisymmetric part if and only if described in the original basis $\mathcal{B}$.

Now we pass to the Clifford algebraic structure by denoting, as usual, the standard Clifford product by juxtaposition. As the bilinear covariants are to be defined in quantum Clifford algebra for the Minkowski spacetime, we restrict our analysis to $V=\mathbb{R}^{1,3}$. Given $\psi\in\cl(V,g)$, the $B$-products between homogeneous  multivectors and arbitrary multivectors are defined  as follows:\medbreak
\noindent a) $u$ is a 1-vector:\beq
u\pu\psi = u\dot\wedge\psi + u\JJgg\psi = u\wedge\psi + u\JJAA\psi + \JJgg\psi =  u\psi + u\JJAA \psi
\eeq
b) $uv$ is a 2-vector:
\beq
(uv)\pu\psi &=& uv\psi + u(v\JJAA\psi) - v\wedge (u\JJAA\psi) + u\JJAA(v\JJBB\psi)
\eeq
c) $uvw$ is a 3-vector:
\beq
(uvw)\pu\psi&=& uvw\psi + uv(w\JJAA\psi) -uw(u\JJgg\psi) + w\wedge(u\JJgg(v\JJAA\psi))
+u(v\JJAA(w\JJBB\psi)) \\&&
+ v\wedge w\wedge (u\JJAA\psi)-v\wedge(u\JJga(w\JJgg\psi))+u\JJAA((v\JJgg w)\psi)-v\wedge(u\JJAA(w\JJBB\psi)) \nonumber\\&&- (w\JJAA u)v\psi -(v\JJAA w)u\psi-v\wedge(u\JJAA w)\psi \,.\nonumber
\eeq
In Eq.~(\ref{c4}) we used the minimal ideal generated by the idempotent
\[
f=\frac{1}{4}(1+\upgamma_0)(1+\iota\upgamma_1\upgamma_2)=\frac{1}{4}(1+2\upgamma_0+\iota\upgamma_0\upgamma_1\upgamma_2)\,.
\]
In $\cl(V,B)$ all formalism is recovered when we analogously consider a left ideal $\cl(V,B)f_B$ where 
\beq
\label{fb} 
f_B=\frac{1}{4}(1+\upgamma_0)\pu(1+\iota\upgamma_1\pu\upgamma_2)\label{012}\,.
\eeq
All results in the last section for $\cl(V,g)$ are obtained \emph{mutatis mutandis}, just by changing the standard Clifford product
$\upgamma_\mu\upgamma_\nu$ by the quantum one:
\beq
\label{q1}
\upgamma_\mu\upgamma_\nu\mapsto\upgamma_\mu\underset{B}{}\upgamma_\nu=\upgamma_\mu\upgamma_\nu + A_{\mu\nu}\,.
\eeq
The last expression is the prominent essence of transliterating $\cl(V, B)$ to $\cl(V,g)$. In $\cl(V,B)$ we have the left ideal $\cl(V,B)f_B$ generated by (\ref{fb}) which can be expanded as  
\begin{eqnarray}
f_B &= \frac{1}{4}(1+\upgamma_0)(1+\iota\upgamma_1\upgamma_2) + \frac{\iota}{4}(A_{12}\upgamma_0 + A_{20}\upgamma_1 + A_{01}\upgamma_2) = f +f(A)\,,
\label{f11}
\end{eqnarray}
where ${f}(A) = \frac{i}{4}(A_{12}\upgamma_0 + A_{20}\upgamma_1 + A_{01}\upgamma_2)$. Our formalism is independent of representation. Just to have some intuition, in the Dirac representation, the idempotent 
$f_B$ in~(\ref{fb}) reads $f = {\rm diag}\,(1,0,0,0)$, and by substituting the expression  
\beq
\upgamma_0\underset{B}{}\upgamma_1\underset{B}{}\upgamma_2 = \upgamma_0\upgamma_1\upgamma_2 + A_{01}\upgamma_2 + A_{20}\upgamma_1 + A_{12}\upgamma_0,\label{012}
\eeq
(which is exactly Eq.~(\ref{dott}))  in~(\ref{f11}), we obtain 
\beq
f_B =  
\begin{pmatrix}1&0&0&0\\0&0&0&0\\
0&0&0&0\\0&0&0&0\end{pmatrix} + 
\begin{pmatrix}\iota A_{12}&0&0&-\iota A_{20}-A_{01}\\0&\iota A_{12}&-\iota A_{20}+A_{01}&0\\
0&\iota A_{20}+A_{01}&-\iota A_{12}&0\\\iota A_{20}-A_{01}&0&0&-\iota A_{12}
\end{pmatrix}\,.
\eeq
When $A_{\mu\nu} = 0$, the latter implies that the bilinear form $B$ is symmetric ($B = g$) and the standard spinor formalism is recovered. Let us denote by $\clb$ the Clifford algebra $\cl(V,B)$, where $V=\mathbb{R}^4$. We adopt a new basis $\{{{\rm e}_\mu}_B\}$ for $\cl(V,B)$ via the $B$-product, which is related to the basis $\{{\rm e}_\mu\}\subset\cl(V,g)$ by 
$${{\rm e}_\mu}_B = \exp(A/2)\; {\rm e}_\mu\; \exp(-A/2)$$~\cite{arbi}, what further shows in a different manner  that $\cl(V,g)$  and $\cl(V,B)$ are isomorphic. 

An arbitrary element of $\psi_B\in \clb$ is written as
\begin{eqnarray}
\psi_B  = h+h^{\mu}\upgamma_{\mu}+h^{\mu\nu}\upgamma_{\mu}\pu\upgamma_\nu+h^{\mu\nu\sigma}\upgamma_{\mu}\pu\upgamma_\nu\pu\upgamma_{\sigma}+p\upgamma_{0}\pu\upgamma_{1}\pu\upgamma_{2}\pu\upgamma_{3}. 
\label{spinorq}
\end{eqnarray}%
Using Eqs.~(\ref{q1}, \ref{012}), we can write~(\ref{spinorq}) as 
\begin{eqnarray}
\psi_B       &=& \psi + h^{\mu\nu}A_{\mu\nu} + h^{\mu\nu\rho}(A_{\mu\nu}\upgamma_\rho + A_{\rho\mu}\upgamma_\nu + A_{\nu\rho}\upgamma_\mu) + \epsilon^{\mu\nu\rho\sigma}A_{\mu\nu}(\upgamma_{\rho}\upgamma_{\sigma} + 
          A_{\rho\sigma})\label{psib}
\end{eqnarray}
where $\psi$ is an arbitrary element in the standard Clifford algebra $\cle$ given by (\ref{spinor}).  Hereon we shall shorten Eq.~(\ref{psib}) to
\begin{eqnarray}
\psi_B = \psi +\psi(A)\,.
\label{psi11}
\end{eqnarray}
The latter display shows that an arbitrary element of $\clb$ can be written as a sum of an arbitrary element $\psi$ of $\cle$ and an $A$-dependent element $\psi(A)$ of $\cle$. Here, 
$\psi(A)\in\bigwedge^0(V)\oplus\bigwedge^1(V)\oplus\bigwedge^2(V)$ and  
\begin{eqnarray}
\psi(A) = (h^{\mu\nu}A_{\mu\nu} + \epsilon^{\mu\nu\rho\sigma}A_{\mu\nu}A_{\rho\sigma})+ h^{\mu\nu\rho}(A_{\mu\nu}\upgamma_\rho + 
          A_{\rho\mu}\upgamma_\nu + A_{\nu\rho}\upgamma_\mu) + \epsilon^{\mu\nu\rho\sigma}A_{\mu\nu}\upgamma_{\rho}\upgamma_{\sigma}
\label{pisa}
\end{eqnarray} 
where $\epsilon^{\mu\nu\rho\sigma}$ denotes the Levi-Civita symbol.

An algebraic $B$-spinor is defined to be an element of the ideal $(\CC\otimes\clb)\pu f_B$. Using Eqs.~(\ref{f11},\ref{psi11}), an arbitrary algebraic $B$-spinor can be written as\footnote{Recall that in the Clifford algebra $\cle$ the product is denoted by juxtaposition.}:
\beq
(\psi_B)\pu(f_B) &=& (\psi + \psi(A))\pu (f+f(A))\nonumber\\
&=& (\psi)\pu f + \psi(A)\pu f + (\psi)\pu f(A) + (\psi(A))\pu f(A)\label{psidis}
\eeq
Hereupon we shall denote by  $s=(h^{\mu\nu}A_{\mu\nu} + \epsilon^{\mu\nu\rho\sigma}A_{\mu\nu}A_{\rho\sigma})$ the scalar part of $\psi(A)$ (see~(\ref{pisa})). Each term in Eq.~(\ref{psidis}) is explicitly shown in the appendix.

\section{Spinor Fields Classification in Quantum Clifford Algebras}
The main goal of this section is to provide some physical insight into the mathematical formalism presented thus far and describe the correspondence between spinor fields in the spacetime and the quantum spinor fields, already named \emph{$B$-spinor fields}.
 
As in the orthogonal Clifford algebraic formalism, the Lounesto's  quantum spinor field classification is given by the following spinor field classes~\cite{lou2}, where in the first three classes it is implicit that $\mathbf{J}_B, \mathbf{K}_B, \mathbf{S}_B \neq 0$:

\begin{itemize}
\item[$1_B$)] $\upsigma_B\neq0,\;\;\; \upomega _B\neq0$.

\item[$2_B$)] $\upsigma_B\neq0,\;\;\; \upomega _B= 0$.\label{dirac1}

\item[$3_B$)] $\upsigma_B= 0, \;\;\;\upomega _B\neq0$.\label{dirac2}

\item[$4_B$)] $\upsigma_B= 0 = \upomega _B, \;\;\;\mathbf{K}_B\neq0,\;\;\; \mathbf{S}_B\neq0$.%
\label{tipo4}

\item[$5_B$)] $\upsigma_B= 0 = \upomega _B, \;\;\;\mathbf{K}_B= 0, \;\;\;\mathbf{S}_B\neq0$.%
\label{type-(5)1}

\item[$6_B$)] $\upsigma_B= 0 = \upomega _B, \;\;\; \mathbf{K}_B\neq0, \;\;\; \mathbf{S}_B = 0$%
.
\end{itemize}
It is always possible to write:
\beq
\upsigma_B &=& \upsigma + \upsigma(A),\qquad\qquad\qquad\qquad
{\mathbf J}_B = {\mathbf J} + {\mathbf J}(A),\nonumber\\
{\mathbf S}_B &=& {\mathbf S} + {\mathbf S}(A),\qquad\qquad\qquad\qquad
{\mathbf K}_B = {\mathbf K} + {\mathbf K}(A), \quad\;{\rm and}\quad\;\;\;\nonumber\\
\upomega _B &=& \upomega  + \upomega (A)\,. \nonumber
\eeq
In general, since we assume the antisymmetric part of the bilinear form $A\neq 0$ (otherwise there is nothing new to prove), it follows that all the  $A$-dependent quantities $\upsigma(A), {\mathbf J}(A), {\mathbf S}(A), {\mathbf K}(A)$, and $\upomega (A)$ do not equal zero.
The expressions for such $A$-independent terms are developed in the appendix.

There is an immediate duality between the spinor fields in the Lounesto classification and the quantum spinor fields that are distributed in the six classes $1_B) - 6_B)$ above:
\begin{itemize}
\item[$1_B$)] $\upsigma_B\neq0,\;\;\; \upomega _B\neq0$.
As $\upsigma_B\neq0$ and $\upsigma_B = \upsigma + \upsigma(A)$, we have some possibilities, depending whether 
$\upsigma$ either does or does not equal zero, as well as $\upomega $:
\begin{enumerate}
\item[$i)$] $\upsigma = 0 = \upomega $.  This case corresponds to the  type-(4), type-(5), and type-(6) spinor fields --- respectively flag-dipoles, flagpoles, and Weyl. Such possibility is obviously compatible to $\upsigma_B\neq0,\;\;\; \upomega _B\neq0$.
\item[$ii)$] $\upsigma = 0$ and $\upomega \neq 0$. This case corresponds to the  type-(3) Dirac spinor fields. 
The condition $\upsigma = 0$ is compatible to $\upsigma_B\neq0$, but as $\upomega \neq 0$, the additional condition
$\upomega _B = \upomega  + \upomega (A) \neq 0$ must be imposed. Equivalently, $0\neq \upomega  \neq \upomega (A)$.
\item[$iii)$] $\upsigma \neq 0$ and $\upomega  = 0$. This case corresponds to the  type-(2) Dirac spinor fields. 
The condition $\upomega  = 0$ is compatible to $\upomega _B\neq0$, but as $\upsigma\neq 0$, the additional condition
$\upsigma_B = \upsigma + \upsigma(A) \neq 0$ must be imposed. Equivalently, $0\neq \upsigma \neq \upsigma(A)$.
\item[$iv)$] $\upsigma \neq 0$ and $\upomega \neq 0$. This case corresponds to the  type-(1) Dirac spinor fields. 
Here both the conditions  $0\neq \upomega  \neq \upomega (A)$ and $0\neq \upsigma \neq \upsigma(A)$ must be imposed.
 \end{enumerate}
All the conditions imposed heretofore must hold in order that the $B$-spinor field be a representing spinor field in class $1_B)$.

\item[$2_B$)] $\upsigma_B\neq0,\;\;\; \upomega _B= 0$.\label{dirac1b}
Although the condition $\upsigma_B\neq0$ is compatible to both the possibilities $\upsigma = 0$ and $\upsigma\neq 0$ (clearly the condition $\upsigma\neq0$ is compatible to $\upsigma_B\neq0$ if $\upsigma\neq -\upsigma(A)$), the condition $\upomega _B= 0$ implies that $\upomega  = - \upomega (A)$, which does not equal zero. To summarize:
\begin{enumerate}
\item[$i)$] $\upsigma = 0$ and $\upomega \neq 0$. This case corresponds to the  type-(3) Dirac spinor fields. 
The condition $\upsigma = 0$ is compatible to $\upsigma_B\neq0$, but as $\upomega \neq 0$, the additional condition
$\upomega _B = \upomega  + \upomega (A) \neq 0$ must be imposed. Equivalently, $0\neq \upomega  \neq \upomega (A)$.
\item[$ii)$] $\upsigma \neq 0$ and $\upomega \neq 0$. This case corresponds to the  type-(1) Dirac spinor fields. 
Here both the conditions  $0\neq \upomega  \neq \upomega (A)$ and $0\neq \upsigma \neq \upsigma(A)$ must be imposed.
 \end{enumerate}
\item[$3_B$)] $\upsigma_B= 0, \;\;\;\upomega _B\neq0$.\label{dirac2b}
Despite the condition $\upomega _B\neq0$ is compatible to both the possibilities $\upomega  = 0$ and $\upomega \neq 0$ (clearly the condition $\upomega \neq0$ is compatible to $\upomega _B\neq0$ if $\upomega \neq -\upomega (A)$), the condition $\upsigma_B= 0$ implies that $\upsigma = - \upsigma(A)$, which does not equal zero. To summarize:
\begin{enumerate}
\item[$i)$] $\upomega = 0$ and $\upsigma\neq 0$. This case corresponds to the  type-(2) Dirac spinor fields. 
The condition $\upomega  = 0$ is compatible to $\upomega _B\neq0$, but as $\upsigma\neq 0$, the additional condition
$\upsigma_B = \upsigma + \upsigma(A) \neq 0$ must be imposed. Equivalently, $0\neq \upsigma \neq \upsigma(A)$.
\item[$ii)$] $\upsigma \neq 0$ and $\upomega \neq 0$. This case corresponds to the  type-(1) Dirac spinor fields. 
Here both the conditions  $0\neq \upomega  \neq \upomega (A)$ and $0\neq \upsigma \neq \upsigma(A)$ must be imposed.
 \end{enumerate}

\item[$4_B$)] $\upsigma_B= 0 = \upomega _B, \;\;\;\mathbf{K}_B\neq0,\;\;\; \mathbf{S}_B\neq0$.%
\label{tipo4b}

\item[$5_B$)] $\upsigma_B= 0 = \upomega _B, \;\;\;\mathbf{K}_B= 0, \;\;\;\mathbf{S}_B\neq0$.%
\label{type-(5)1b}

\item[$6_B$)] $\upsigma_B= 0 = \upomega _B, \;\;\; \mathbf{K}_B\neq0, \;\;\; \mathbf{S}_B = 0$%
.
\end{itemize}
All the quantum spinor fields $4_B$), $5_B$), and $6_B$) are defined by the condition $\upsigma_B= 0 = \upomega _B$.
It implies that $\upsigma = -\upsigma(A) (\neq 0)$, and that $\upomega  = -\upomega (A) (\neq 0)$. It means that all the singular $B$-spinor fields correspond to the  type-(1) Dirac spinor fields. 

All that it is above discussed can be encrypted 
in Table I.
 \medbreak
 \begin{table}
\centering
\begin{tabular}{||r|r||r|r||}
\hline\hline
&Quantum Spinor Fields&Spinor Fields&\\\hline\hline
 type-($1_B$)& $B$-Dirac&Dirac&type-(1)\\
  &&Dirac&type-(2)\\
  & &Dirac&type-(3)\\
&  &Flag-dipoles&type-(4)\\
&  &Flagpole (Elko, Majorana, \ldots) &type-(5)\\
&  &Dipole (Weyl, \ldots) &type-(6)\\\hline
type-($2_B$)& $B$-Dirac&Dirac&type-(3)\\
  & &Dirac&type-(1)\\\hline
type-($3_B$)& $B$-Dirac&Dirac&type-(2)\\
  & &Dirac&type-(1)\\\hline
  type-($4_B$)& $B$-flag-dipole&Dirac&type-(1)\\\hline 
  type-($5_B$) & $B$-flagpole&Dirac&type-(1)\\\hline
 type-($6_B$)& $B$-dipole&Dirac&type-(1)\\\hline
  
\hline\hline
\end{tabular}\vspace{0.5mm}
\caption{Correspondence among the standard spinor fields and the (quantum) $B$-spinor fields under Lounesto spinor field classification.}
\label{table1}
\end{table}

\section{Concluding Remarks and Outlook}
We have shown that the introduction of an  antisymmetric part into the spacetime metric drastically alters the structure of spinor fields and their classification based on the bilinear covariants. 
All the results deriving a duality between standard and quantum spinor fields are condensed in Table I.  The distribution of intrinsic angular momentum, formerly a legitimate bivector in the standard Clifford algebra $\cl(V,g)$, is now the direct sum of a bivector and a scalar which shows different $\mathbb{Z}$-grading induced by the arbitrary bilinear form in the underlying vector space. Furthermore, the direction of the electron spin ${\mathbf K}$ is now a paravector (the sum of a scalar and a vector), which is not a homogeneous Clifford element. Indeed, in $\cl(V,B)$ it is homogeneous, but in $\cl(V,g)$ it is a paravector. It is useful to look at the Clifford algebra $\cl(V,B)$  in the $\bigwedge(V)$ basis and not in the canonical $\dot{\bigwedge}(V)$. As previously already asserted, in order to perform experiments and phenomenology concerning the $B$-bilinear covariants, the basis for $\bigwedge(V)$, and not for $\dot\bigwedge(V)$ should be taken into account, by considering an independent coordinatization  in a projective setup. In the quantum field theory, the self duality of this coordinatization implies precisely the Feynman propagator. Bilinear covariants exhibit, however, more structure, e.g., the dualities. These do change and one can use this to obtain Hecke algebras, Yang-Baxter solutions~\cite{aca}, some issues that are far beyond the scope here which was to provide a complete spinor field classification according to the bilinear covariants in a quantum Clifford algebra for the Minkowski spacetime.

The classification encoded in Table I present a straightforward way to probe the bilinear form that endows the spacetime. Moreover, it reveals the duality between the standard and the quantum spinor fields, under Lounesto's spinor field classification, by proposing 
which classes of spinor fields can be connected 
by the introduction of a antisymmetric part in the spacetime metric.  Besides the more general mathematical aspects, both field theory and gravitation comply not to fix ab initio 
the vector space to be endowed with a symmetric bilinear form.

\section{Acknowledgement}
RdR thanks CNPq Grants No. 473326/2013-2 and No. 303027/2012-6 for partial financial support, also being \emph{Bolsista da CAPES Proc. n$^{o}$} 10942/13-0. IG is grateful to FAPESP grants 2012/07710-0 (BEPE) and 2011/04918-7 and authors are thankful to Prof. Bertfried Fauser for suggestions. 
\appendix

\section{Additional Terms in the Quantum Spinor Fields} 
Eq.~(\ref{psidis}) is given by
\begin{gather}
(\psi_B)\pu(f_B) =  (\psi)\pu f + \psi(A)\pu f + (\psi)\pu f(A) + (\psi(A))\pu f(A)\label{psidis1}
\end{gather} 
and one can see that $(\psi)\pu f$ is the classical spinor field in~(\ref{c4}). The additional terms in  the  equation above represent correction terms, provided by:
\medbreak
a) The term $\boxed{-4\iota(\psi(A))\pu f(A)}$ is given by
{\footnotesize{\beq
&&\big\{p\left[h^{013}(A_{01}(A_{01}A_{32}+A_{20}A_{31}+A_{12}A_{30})+A_{12}A_{13}+A_{03}A_{20})\right.\nonumber\\&&\left.
+h^{023}(A_{02}(A_{01}A_{32}+A_{20}A_{31}+A_{12}A_{30}+2A_{30})+A_{12}A_{13})\right.\nonumber\\
&&+\left.h^{123}(A_{12}(A_{01}A_{32}+A_{20}A_{31}+A_{12}A_{30})-A_{23}A_{20}-A_{31}A_{01})+
h^{012}(A_{10}A_{01}+A_{20}A_{02}+A_{12}A_{12})\right]\big\}\nonumber\\
&+&\upgamma_0\left[p\left(A_{13}A_{01}-A_{23}A_{20}+2A_{12}A_{12}A_{13}+A_{23}A_{20}A_{12}+A_{23}A_{01}A_{12}\right)+sA_{12}\right]
\nonumber\\
&+&\upgamma_1\left[p\left(A_{12}A_{13}-A_{12}A_{23}-A_{03}A_{01}+A_{01}A_{20}A_{32}+A_{01}A_{20}A_{13}+A_{02}A_{12}A_{03}+A_{03}A_{12}A_{21}+A_{23}A_{01}A_{10}\right)\right.\nonumber\\&&\left.\qquad\qquad\qquad\qquad+sA_{01}\right]
\nonumber\\
&+&\upgamma_2\left[p\left(A_{03}A_{20}+A_{01}A_{01}A_{32}+A_{13}A_{01}A_{02}+A_{13}A_{20}A_{02}\right)+sA_{02}\right]
\nonumber\\
&+& \upgamma_3\left[p\left(A_{01}A_{01}+A_{02}A_{02}+A_{02}A_{12}A_{13}+A_{12}A_{20}A_{10}+A_{02}A_{20}A_{12}+A_{01}A_{12}A_{13}\right)\right]
\nonumber\\
&+&\upgamma_{01}\left[p\left(h^{013}(A_{13}A_{20}+A_{21}A_{30})+h^{023}(A_{03}A_{12}+A_{13}A_{20})-h^{123}A_{23}A_{12}\right)\right]
\nonumber\\
&+&\upgamma_{02}\left[p\left(h^{013}A_{13}A_{01}+h^{023}A_{13}A_{01}+h^{123}A_{13}A_{21}\right)\right]
+\upgamma_{03}\left[p\left(h^{013}A_{01}A_{12}+h^{023}A_{02}A_{12}+h^{123}A_{12}A_{21}\right)\right]
\nonumber\\
&+&\upgamma_{12}\left[p\left(h^{013}A_{01}A_{30}+h^{023}A_{30}A_{01}+h^{123}(A_{13}A_{20}+A_{23}A_{01})\right)\right]
\nonumber\\
&+&\upgamma_{31}\left[p\left(h^{013}A_{01}A_{20}+h^{023}A_{01}A_{20}+h^{123}A_{12}A_{20}\right)\right]
+\upgamma_{23}\left[p\left(h^{013}A_{01}A_{10}+h^{023}A_{01}A_{20}+h^{123}A_{12}A_{10}\right)\right]
\nonumber\\
&+& \upgamma_{023}A_{12}A_{01} + \upgamma_{031}(A_{01}A_{12}+A_{12}A_{20}+A_{23}A_{01})+\upgamma_{012}(A_{03}A_{12}+A_{13}A_{20}))
\nonumber\eeq}}
\medbreak
b) The term $\boxed{-4\iota(\psi(A))\pu f}$ is given by
{\footnotesize{\beq
&&\left[h^{023}(A_{03}+A_{02})+h^{123}A_{23}+h^{012}A_{01}+h^3A_{01}A_{32}+h^3(A_{20}A_{31}+A_{03}A_{21})+h^0A_{12}+h^2A_{01}+h^1A_{02}\right]\nonumber\\
&+&\upgamma_0\left[h^{01}\left(A_{01}+A_{20})A_{12}+A_{20}\right)+h^{02}(A_{10}+A_{12}A_{20})+h^{03}(A_{01}{32}+A_{20}A_{32}+A_{12}A_{30}) + bA_{12}\right.\nonumber\\&&\left.+ p\left(A_{12}+A_{20}A_{13}+A_{10}(A_{23}+A_{13})+A_{30}A_{12}\right)\right]
\nonumber\\
&+&\upgamma_1\left[h^{01}\!\left(A_{20}(A_{01}\!+\!A_{02}) \!+\!A_{12}(A_{10}\!+\!1)\right)\!+\!bA_{20}\!+\!h^{12}((A_{12}\!+\!A_{02})\!A_{20}\!-\!A_{01})A_{01}\!+\!h^{13}(A_{32}(A_{01}+A_{20})\!+\!A_{12}A_{30})\right]
\nonumber\\
&+&\upgamma_2\left[h^{02}\!A_{21} \!+\! h^{12}A_{20} \!+\! h^{23}\!\left(A_{01}A_{32}\!+\!A_{12}A_{31}\!+\!A_{20}A_{32}\!+\!bA_{01}\!+\!p(A_{12}A_{31}+2A_{02}(A_{01}A_{31} +A_{21}A_{30})\!+\!A_{20}A_{20}) \right)\right]
\nonumber\\
&+& \upgamma_3\left[p\left(-A_{01}A_{01}-A_{02}A_{02}+A_{12}A_{12})+h^{03}A_{21})+h^{13}(A_{01}A_{21}+A_{20})+h^{23}A_{12}(A_{02}+A_{01}A_{13})\right)\right]
\nonumber\\
&+&\upgamma_{01}\left[\left(h^{013}(A_{31}A_{20}+A_{12}A_{30}+A_{32}A_{01})-h^{013}(A_{03}A_{12}+2A_{13}A_{20})+h^{123}A_{23}(A_{12}+A_{02})+h^0A_{20}+h^1A_{21}\right)\right]
\nonumber\\
&+&\upgamma_{02}\left[\left(h^{013}A_{31}A_{01}+h^{023}(A_{32}A_{01}+A_{20}A_{31}+A_{30}A_{12})+h^{123}A_{32}A_{01}+h^{012}A_{21}A_{01}+(h^0+h^2)A_{01}\right)\right]\nonumber\\&&
+\upgamma_{03}\left[2h^{012}A_{20}+h^3A_{12}\right]
\nonumber\\
&+&\upgamma_{12}\left[(h^1A_{01}+h^2A_{02}+h^{012}A_{12} + h^{023}(A_{13}A_{01}+A_{03}A_{20})+h^{123}A_{30}A_{01}+h^{123}(A_{30}A_{12}+A_{20}A_{31}))\right]
\nonumber\\
&&-\upgamma_{31}\left[\left(h^{013}A_{01}A_{20}+h^{012}A_{01}A_{20}+h^{123}A_{01}\right)\right]
+\upgamma_{23}\left[\left(h^{023}A_{12}+h^{123}(A_{10}A_{12}+A_{20}+A_{02}A_{01}) + h^{3}A_{10}\right)\right]
\nonumber\\
&+& \upgamma_{023}(h^{03}A_{10} + h^{23}A_{12} + p(A_{20}A_{12} + A_{02} + A_{21}A_{10}))+ \upgamma_{013}[p(2A_{20}A_{02}+A_{01}) +h^{03}A_{02}+h^{13}A_{12}]\nonumber\\&&+\upgamma_{012}[h^{01}A_{01}+h^{02}A_{02}+h^{12}A_{12} + p(A_{02}A_{23}+A_{13}A_{20}+A_{21}A_{20})]\nonumber\\
&+&\upgamma_{123}[p(A_{21}+A_{20}A_{20})+h^{23}A_{20}+h^{13}A_{10}]+\upgamma_{0123}(h^{013}A_{10}+h^{023}A_{20}+h^{123}A_{12})\,.
\nonumber\eeq}}
c) The term $\boxed{-4\iota\psi\pu f}$ is given by
{\footnotesize{\beq
&&\left[h^{012}(A_{21}A_{12}-1)+h^{013}A_{31}A_{02}+h^{023}A_{31}+h^{123}(A_{30}-A_{32}A_{20})+p\left(A_{02}(A_{23}+A_{13}+A_{01}A_{12})+A_{01}(A_{32}+A_{31})\right.\right.\nonumber\\&&\left.\left.+(A_{21}-1)A_{03})\right)\right]\nonumber\\
&+&\upgamma_0\left[h^{01}\left(A_{21}A_{01}+A_{02}+A_{10})\right)+h^{02}(A_{21}A_{20}+A_{01}+A_{20})+h^{03}(A_{02}{31}+A_{23}A_{01}+A_{30}) +h^{31}\left(A_{21}A_{31}+A_{23})\right)\right.\nonumber\\&&\left.+ p\left(A_{30}A_{20}+A_{32}A_{01}+ A_{21}A_{20}A_{13}+A_{30}A_{21}A_{12}   + A_{30}A_{20}+A_{32}A_{01}+ A_{21}A_{12}(A_{02}+A_{30})+h^{023}A_{23}A_{21}\right) \right.\nonumber\\&&\left.+h^{12}(A_{21}A_{12}-1)+h^{23}(A_{31}-A_{32}A_{21})\right]
\nonumber\\
&+&\upgamma_1\left[h^{01}\left(A_{20}A_{10}+A_{12}+2)\right)+h^{02}(A_{20}A_{02}-1)+h^{03}(A_{32}-A_{30}A_{02}) +h^{31}\left(A_{01}A_{32}+A_{12}A_{30}+A_{30})\right)\right.\nonumber\\&&\left.+h^{12}(A_{20}A_{12}+A_{01}+A_{02})+h^{23}(A_{30}+A_{32}A_{20}) + p\left(A_{32}A_{01}A_{02}+ A_{30}A_{20}A_{12}\right)\right]\nonumber\\
&+&\upgamma_2\left[h^{01}\left(A_{10}A_{01}-1\right)+h^{02}(A_{20}A_{01}+A_{12}+2)+h^{03}(A_{30}A_{01}+A_{13}) +h^{31}\left(A_{10}A_{31}+A_{30})\right)\right.\nonumber\\&&\left.+h^{12}(A_{10}A_{12}+A_{02})+h^{23}(A_{30}A_{21}+A_{31}A_{20}) + p\left(A_{01}(A_{12}A_{30}+A_{32}A_{10})+ A_{23}\right)\right]
\nonumber\\
&+& \upgamma_3\left[h^{123}A_{23}+2h^{03}+2h^{13}A_{10}+2h^{23}A_{20}+pA_{20}\right] +\upgamma_{03}\left[h^{123}A_{32}\right]
\nonumber\\
&+&\upgamma_{01}\left[\left(h^{013}(A_{03}A_{12})+h^{012}A_{03}A_{12}+h^{123}(A_{13}+A_{23}A_{12})+h^{023}A_{30}+p(A_{20}(A_{32}A_{10}+A_{31}A_{20})+A_{31}+A_{12}A_{23})\right)\right]
\nonumber\\
&+&\upgamma_{02}\left[h^{013}A_{03}+h^{012}A_{10}+h^{123}A_{32}+h^{023}(A_{30}A_{21}+A_{23}A_{01}+A_{13}A_{20})+p(A_{10}(A_{32}A_{10}+A_{31}A_{20})+A_{23})\right]
\nonumber\\
&+&\upgamma_{12}\left[h^{012}(A_{21}+A_{02}A_{01}) + h^{023}A_{32}+h^{123}A_{30}(A_{21}+A_{10})+h^{013}(A_{31}+A_{03}A_{10})+p(A_{30}A_{12}+A_{23}A_{10}+A_{13}A_{20})\right]
\nonumber\\
&+&\upgamma_{31}\left[h^{013}A_{12}+pA_{02}A_{21}\right]
+\upgamma_{012}[h^{01}A_{01}+h^{02}A_{02}+h^{12}A_{12} + p(A_{02}A_{23}+A_{13}A_{20}+A_{21}A_{20})]\nonumber\\
&+& \upgamma_{023}(h^{03}A_{10} + h^{31}+h^{23}A_{12} + p(A_{10}A_{12} + A_{02} + A_{21}A_{01}))
+ \upgamma_{013}[p(A_{20}A_{21}+A_{10}) +h^{03}A_{20}-h^{23}+h^{31}A_{12}]\nonumber\\&+&\upgamma_{123}[p(A_{21}(A_{30}+A_{32})+A_{10}A_{32})+h^{23}A_{02}+h^{31}A_{10}]+\upgamma_{0123}(h^{023}A_{02}+h^{123}A_{12})\,.
\nonumber\eeq}}
d) The term $\boxed{-4\iota\psi\pu f(A)}$ is given by
{\footnotesize{\beq
&&\left[h^{0}A_{12}+h^2A_{10}+h^1A_{02}\right]\nonumber\\
&+&\upgamma_0\left[h^{01}\left(A_{12}A_{01}+A_{02}A_{21}+A_{20})\right)+h^{02}(A_{10}+A_{12}A_{20})+h^{03}(A_{01}A_{32}+A_{23}A_{02}+A_{12}A_{30}) +bA_{12} + \right.\nonumber\\&&\left.pA_{12}\left(A_{20}A_{13}+A_{32}A_{01}+ A_{30}A_{12}+A_{10}A_{13}\right)\right]
\nonumber\\
&+&\upgamma_1\left[h^{01}(A_{20}(A_{01}+A_{02})+A_{12}(A_{01}+1))+h^{12}(A_{10}+A_{02}A_{20}+A_{12}A_{20})+h^{13}(A_{01}A_{32}+A_{23}A_{02}+A_{12}A_{30})+bA_{20} \right.\nonumber\\&&\left.+ p\left(A_{12}A_{23}+A_{10}A_{30}+ A_{02}A_{01}A_{23}\right)\right]
\nonumber\\
&+&\upgamma_2\left[h^{02}A_{21}+h^{12}A_{20} +h^{23}(A_{12}A_{30}+A_{32}A_{01}+A_{20}A_{32})+p(A_{20}(2A_{10}A_{31}+2A_{12}A_{30}+A_{20})+A_{12}A_{31})\right]
\nonumber\\
&+& \upgamma_3\left[A_{20}A_{02}+A_{10}A_{01}+A_{12}A_{12}+h^{03}A_{21})+h^{13}(A_{20} +A_{01}A_{21})+h^{23}A_{01}\right]+\upgamma_{03}\left[h^{012}A_{20}\right]
\nonumber\\
&+&\upgamma_{01}\left[\left(h^{013}(A_{03}A_{12}+A_{31}A_{20})+h^{123}A_{23}(A_{12}+A_{02})+h^{012}(A_{10}+A_{20}A_{21})+h^{013}A_{32}A_{01}+h^0A_{20}+h^1A_{21}\right)\right]
\nonumber\\
&+&\upgamma_{02}\left[\left(h^{013}A_{01}A_{31}+h^{123}A_{32}A_{01}+h^{012}(A_{20}+A_{21}A_{01})+h^{023}(A_{23}A_{01}+A_{20}A_{31}+A_{30}A_{12})+h^0A_{01}+h^2A_{01}\right)\right]
\nonumber\\
&+&\upgamma_{12}\left[\left(h^{013}A_{01}A_{31}+h^{123}(A_{20}A_{31}+A_{30}A_{12}+A_{32}A_{01})+h^{023}(A_{13}A_{01}+A_{03}A_{20})+h^1A_{01}+h^2A_{02}\right)\right]
\nonumber\\
&+&\upgamma_{13}\left[h^{013}A_{12}+h^{123}A_{01} +h^{013}A_{02}+h^3A_{02}\right]+\upgamma_{23}\left[h^{123}\left(A_{01}(A_{02}+A_{21})+A_{20}\right)+h^{023}A_{12}\right]\,.
\nonumber
\eeq}}




\footnotesize
\begin{thebibliography}{99}
\footnotesize
\bibitem{Hawking}
S.~W.~Hawking, \emph{The Unpredictability of Quantum Gravity}, Commun.\ Math.\ Phys.\  \textbf{87} (1982) 395.

\bibitem{chevalley} 
C.~Chevalley, ``The Algebraic Theory of Spinors'', Columbia Univ. Press, 1954. 

\bibitem{Oziewicz97} 
Z.~Oziewicz, {\it Clifford algebra of multivectors\/}, Adv. Appl. Clifford Algebras \textbf{7} (Suppl.) (1997)  467.

\bibitem{Fsr1} 
B.~Fauser, {\it On an easy transition from operator dynamics to generating functionals by Clifford algebras}, J. Math. Phys. \textbf{39} (1998) 4928. 

\bibitem{Fsr2} 
B.~Fauser, {\it Vertex normal ordering as a consequence of nonsymmetric bilinear forms in Clifford algebras\/}, J. Math. Phys. \textbf{37} (1996) 72 [{\tt arXiv:hep-th/9504055}].

\bibitem{Fsr3} 
B.~Fauser, {\it Hecke algebra representations within Clifford geometric algebras of multivectors\/}, J.~Phys.~A: Math. Gen. \textbf{32} (1999) 1919.

\bibitem{ozie} 
Z.~Oziewicz, {\it Clifford algebra for Hecke braid\/}, in ``Clifford Algebras and Spinor Structures'', Special Volume to the Memory of Albert Crumeyrolle. Eds. R.~Ab\l amowicz, 
P.~Lounesto, Kluwer, Dordrecht, 1995, pp. 397-412.

\bibitem{conformal}
R.~da Rocha, A.~E.~Bernardini, J.~Vaz, Jr. \emph{$\kappa$-deformed Poincar\'{e} algebras and quantum Clifford-Hopf algebras},  Int.\ J.\ Geom.\ Meth.\ Mod.\ Phys.\  \textbf{7} (2010) 821 [{\tt arXiv:0801.4647 [math-ph]}].

\bibitem{ber1}
  A.~E.~Bernardini and R.~da Rocha,
  \emph{Obtaining the equation of motion for a fermionic particle in a generalized Lorentz-violating system framework},
  Europhys.\ Lett.\  {\bf 81} (2008) 40010
  [{\tt arXiv:hep-th/0701092}].

\bibitem{rota} 
G.-C.~Rota, J.~A.~Stein, \emph{Plethystic algebras and vector symmetric functions}, Proc. Natl. Acad. Sci. \textbf{91}  (1994) 13062.

\bibitem{gebra}
B.~Fauser, Z.~Oziewicz, \emph{Clifford Hopf gebra for two dimensional space}, Miscellanea Algebraica \textbf{2} (2001) 31 [{\tt arXiv:math/0011263 [math.QA]}].

\bibitem{vacua}  
B.~Fauser, \emph{Clifford geometric parametrization of inequivalent vacua}, Math. Meth. Appl. Sci. \textbf{24} (2001) 885.

\bibitem{moffat}
J.~W.~Moffat, \emph{Nonsymmetric gravitational theory,} Phys.\ Lett.\ B \textbf{355} (1995) 447 [{\tt arXiv:gr-qc/9411006}].

\bibitem{moffat01} J.~W.~Moffat, \emph{Gravitational Theory, Galaxy Rotation Curves and Cosmology without Dark Matter}, JCAP \textbf{05} (2005) 3 [{\tt arXiv:astro-ph/0412195}].

\bibitem{moffat1} 
T.~Janssen, T.~Prokopec, \emph{Vacuum properties of nonsymmetric gravity in de Sitter space}, JCAP \textbf{05} (2007) 010 [{\tt arXiv:gr-qc/0703050}]
\bibitem{moffat2}  T.~Janssen, T.~Prokopec, \emph{Problems and hopes in nonsymmetric gravity} J. Phys. A \textbf{40} (2007) 7067 [{\tt arXiv:gr-qc/0611005}].
\bibitem{moffat3}  Y.~Mao, M.~Tegmark, A.~Guth,  S.~Cabi, \emph{Constraining Torsion with Gravity Probe B}, Phys. Rev. D 
\textbf{76} (2007) 104029 [{\tt arXiv:gr-qc/0608121}].
\bibitem{moffat4}   T.~Prokopec, W.~Valkenburg, \emph{The cosmology of the nonsymmetric theory of gravitation}, Phys. Lett. B \textbf{636} (2006) 1 [{\tt arXiv:astro-ph/0503289}]. 



\bibitem{hammond} R. T. Hammond, \emph{Spin from the non symmetric metric tensor}, Int. J. Mod. Phys. D {\bf 22} (2013) 1342009.

\bibitem{Ghosh:2012jh}
  S.~Ghosh and S.~Shankaranarayanan,
  \emph{5D non-symmetric gravity and geodesic confinement}, 
  Gen.\ Rel.\ Grav.\  {\bf 45} (2013) 1787
  [{\tt arXiv:1210.4361 [gr-qc]}].


\bibitem{alex}
  A.~E.~Bernardini, R.~da Rocha,
  \emph{Dynamical dispersion relation for ELKO dark spinor fields,
  Phys.\ Lett.\ B} {\bf 717} (2012) 238
  [{\tt arXiv:1203.1049 [hep-th]}].

\bibitem{horvath} D. V. Ahluwalia, S. P. Horvath, \emph{Very special relativity as relativity of dark matter: the Elko connection, JHEP} {\bf 11} (2010) 078 	[{\tt arXiv:1008.0436 [hep-ph]}].

\bibitem{2cylee}
  D.~V.~Ahluwalia, C.~-Y.~Lee, D.~Schritt,  T.~F.~Watson,
  \emph{Elko as self-interacting fermionic dark matter with axis of locality}, 
  Phys.\ Lett.\ B {\bf 687} (2010) 248
  [{\tt arXiv:0804.1854 [hep-th]}].


\bibitem{cyleeI}
  D.~V.~Ahluwalia, C.~-Y.~Lee, D.~Schritt,
 \emph{Self-interacting Elko dark matter with an axis of locality}, 
  Phys.\ Rev.\ D {\bf 83} (2011) 065017
  [{\tt arXiv:0911.2947 [hep-ph]}].



 \bibitem{canj}
  K.~E.~Wunderle, R.~Dick,
  \emph{Transformation properties and symmetry behaviour of ELKO spinors},
  Can.\ J.\ Phys.\  {\bf 87} (2009) 909.
  \bibitem{dasi}
R.~da Rocha, J.~G.~Pereira,
 \emph{The Quadratic spinor Lagrangian, axial torsion current, and generalizations},
  Int.\ J.\ Mod.\ Phys.\ D {\bf 16} (2007) 1653
  [{\tt arXiv:gr-qc/0703076 [gr-qc]}].  
  \bibitem{coincidence}
  H.~M.~Sadjadi,
  \emph{On coincidence problem and attractor solutions in ELKO dark energy model},
  Gen.\ Rel.\ Grav.\  {\bf 44} (2012) 2329
  [{\tt arXiv:1109.1961 [gr-qc]}].
 
  
 \bibitem{shankara}
  A.~Basak, J.~R.~Bhatt, S.~Shankaranarayanan, K.~V.~Prasantha Varma,
  \emph{Attractor behaviour in ELKO cosmology},
  \emph{JCAP} {\bf 1304} (2013) 025
  [{\tt arXiv:1212.3445 [astro-ph.CO]}].
 \bibitem{liu}
  Y.~-X.~Liu, X.~-N.~Zhou, K.~Yang, F.~-W.~Chen,
  \emph{Localization of 5D Elko Spinors on Minkowski Branes},
  Phys.\ Rev.\ D {\bf 86} (2012) 064012
  [{\tt arXiv:1107.2506 [hep-th]}].
 \bibitem{Annalen Phys.}
  C.~G.~Boehmer,
  \emph{The Einstein-Cartan-Elko system},
  Annalen Phys.\  {\bf 16} (2007) 38
  [{\tt arXiv:gr-qc/0607088}]. 
 \bibitem{vign}
  S.~Vignolo, L.~Fabbri, R.~Cianci,
 \emph{Dirac spinors in Bianchi-I f(R)-cosmology with torsion},
  J.\ Math.\ Phys.\  {\bf 52} (2011) 112502
  [{\tt arXiv:1106.0414 [gr-qc]}].
 
\bibitem{lucafabbri}
  L.~Fabbri,
 \emph{Metric Solutions in Torsionless Gauge for Vacuum Conformal Gravity},
  J.\ Math.\ Phys.\  {\bf 54} (2013) 062501
  [{\tt arXiv:1104.5002 [gr-qc]}].

\bibitem{elkoh}
  R.~da Rocha and J.~M.~Hoff da Silva,
  \emph{Hawking Radiation from Elko Particles Tunnelling across Black Strings Horizon},
  Europhys.\ Lett.\  {\bf 107} (2014) 50001
  [{\tt arXiv:1408.2402 [hep-th]}].
  
\bibitem{accele} M. Dias, F. de Campos, J. M. Hoff da Silva, \emph{Exploring light Elkos signal at accelerators}, 	{Phys. Lett. B} {\bf 706} (2012) 352 [{\tt 	arXiv:1012.4642 [hep-ph]}].



\bibitem{alve}
  A.~Alves, F.~de Campos, M.~Dias,  J.~M.~Hoff da Silva,
 \emph{Searching for Elko dark matter spinors at the CERN LHC},
  [{\tt arXiv:1401.1127 [hep-ph]}].

\bibitem{riemanncartan}
  R.~da Rocha, L.~Fabbri, J.~M.~Hoff da Silva, R.~T.~Cavalcanti, J.~A.~Silva-Neto,
  \emph{Flag-Dipole Spinor Fields in ESK Gravities}, 
  J.\ Math.\ Phys.\  {\bf 54} (2013) 102505
  [{\tt arXiv:1302.2262 [gr-qc]}].



\bibitem{wess} G. Hess, \emph{Exotic Majorana spinors in (3+1)-dimensional space-times}, J. Math. Phys. {\bf 35} (1994)
4848.

\bibitem{Asselmeyer:1995jm}
  T.~Asselmeyer and R.~Keiper,
  \emph{Topological investigation of the fractionally quantized hall conductivity,}
  Annalen Phys.\  {\bf 4} (1995) 739
  [{\tt arXiv:cond-mat/9508055}].

\bibitem{exotic} R.~da Rocha, A.~E.~Bernardini, J.~M.~Hoff da Silva, \emph{Exotic Dark Spinor Fields}, 
  \emph{JHEP} {\bf 04} (2011) 110
  [{\tt arXiv:1103.4759 [hep-th]}].

\bibitem{Cavalcanti:2014wia}
  R.~T.~Cavalcanti,
  \emph{Looking for the Classification of Singular Spinor Fields Dynamics and other Mass Dimension One Fermions: Characterization of Spinor Fields}, 
  [{\tt arXiv:1408.0720 [hep-th]}].

\bibitem {moro} 
R.~A.~Mosna, W.~A.~Rodrigues,~Jr., \emph{The bundles of algebraic and Dirah-Hestenes spinor fields}, J. Math. Phys. \textbf{45} (2004) 2945 [{\tt arXiv:math-ph/0212033}]

\bibitem{yvon} 
J.~Yvon, \emph{Equations de Dirah-Madelung},  J. Phys. et de Radium \textbf{8} (1940) 18.
 
\bibitem{taka} 
T.~Takabayasi, \emph{ Relativistic hydrodynamics of the Dirac matter}, Theor. Phys. Suppl. \textbf{4} (1957) 1.
 
\bibitem{fierz} 
M.~Fierz, \emph{Zur Fermischen theorie des X-zerfalls}, Z. Phys. \textbf{104} (1937) 553.

\bibitem{lou2}
P.~Lounesto, ``Clifford Algebras and Spinors'', 2$^{\mathrm{nd}}$ ed. Cambridge Univ. Press, Cambridge, 2002.

\bibitem {hol}
P.~R.~Holland, \emph{Relativistic Algebraic Spinors and Quantum Motions in Phase Space}, Found. Phys. \textbf{16} (1986) 708.

\bibitem{cra}
J.~P.~Crawford, \emph{On the Algebra of Dirac Bispinor Densities: Factorization and Inversion Theorems}, J. Math. Phys. \textbf{26} (1985) 1429.
  
\bibitem{elko}  
R.~da Rocha, J.~M.~Hoff da Silva, \emph{ELKO Spinor Fields: Lagrangians for Gravity derived from Supergravity}, Int.\ J.\ Geom.\ Meth.\ Mod.\ Phys.\  \textbf{6} (2009) 461 [{\tt arXiv:0901.0883 [math-ph]}].
 
\bibitem{arbi}
B.~Fauser, R.~Ab\l amowicz, \emph{On the decomposition of Clifford algebras of arbitrary bilinear form}, in \textit{Clifford Algebras and their Applications in Mathematical Physics}, eds. R.~Ab{\l}amowicz and B.~Fauser, Vol.~1: Algebra and Physics, Birkh\"{a}user, Boston, 2000, 341--366 [{\tt arXiv:math/9911180 [math-qa]}].   
  
\bibitem{hecke}
R.~Ab\l amowicz, B.~Fauser, \emph{Hecke algebra representations in ideals generated by q-young Clifford idempotents}, in \textit{Clifford Algebras and their Applications in Mathematical Physics}, R.~Ab{\l}amowicz and B.~Fauser, (eds.), Vol.~1: Algebra and Physics, Birkh\"{a}user, Boston, 2000, 245--268 [{\tt math/9908062 [math-qa]}].
   
\bibitem{aca} 
B.~Fauser, \emph{Quantum Clifford Hopf algebra for quantum field theory},  Adv.\ Appl.\ Clifford Algebras \textbf{13} (2003) 115   [{\tt arXiv:hep-th/0011026}].
\end{thebibliography}
\end{document}